\documentclass[10pt,conference,twocolumn]{IEEEtran}

\usepackage{amsmath}
\usepackage{amssymb}
\usepackage{latexsym}
\usepackage{amsfonts}

\begin{document}
\pagestyle{headings}  % switches on printing of running heads
\vspace{1cm}

%
%\title{On the $k$-error linear complexity of $2^n$-periodic binary
%sequences with fixed linear complexity,\\
\title{Characterization of $2^n$-periodic binary sequences with
fixed 3-error or 4-error linear complexity}

\author{
\authorblockN{Jianqin Zhou$^{1,2}$, Jun Liu$^{1}$}%, Feng La$^{[1]}$, Zemao Zhao$^{[1]}$ and Lin You$^{[1]}$ }
\authorblockA{ Telecommunication School, Hangzhou Dianzi University,
Hangzhou, 310018 China\\  Computer Science School, Anhui Univ. of
Technology, Ma'anshan, 243002 China\\ \ \ zhou9@yahoo.com\\
\ \\
Wanquan Liu\\
Department of Computing, Curtin University, Perth, WA 6102 Australia\\
 w.liu@curtin.edu.au
 }
}
\maketitle              % typeset the title of the contribution

\begin{abstract}
The linear complexity and the $k$-error linear complexity of a
sequence have been used as important security measures for key
stream sequence strength in linear feedback shift register design.
By using the sieve method of combinatorics, the $k$-error linear
complexity distribution of $2^n$-periodic binary sequences is
investigated  based on Games-Chan algorithm.
 First, for $k=2,3$, the complete counting functions on
the $k$-error linear complexity of $2^n$-periodic  binary sequences
with linear complexity less than $2^n$ are characterized. Second,
for $k=3,4$, the complete counting functions on the $k$-error linear
complexity of $2^n$-periodic binary sequences with linear complexity
$2^n$ are presented. Third,  for $k=4,5$, the complete counting
functions on the $k$-error linear complexity of $2^n$-periodic
binary sequences with linear complexity less than $2^n$ are derived.
As a consequence of these results, the counting functions for the
number of $2^n$-periodic binary sequences with the $3$-error linear
complexity  are obtained, and the complete counting functions on the
$4$-error linear complexity of $2^n$-periodic binary sequences are
obvious.


\end{abstract}

\section{Introduction}

%\ \\
The linear complexity of a sequence is defined as the length of the
shortest linear feedback shift register (LFSR) that can generate the
sequence. The concept of linear complexity is very useful in the
study of security of stream ciphers for cryptographic applications
and it has attracted many attentions in cryptographic community
\cite{Ding,Stamp}. In fact, a necessary condition for the security
of a key stream generator in LFSR is that it produces a sequence
with high linear complexity. However, high linear complexity can not
necessarily guarantee that the sequence is safe since the linear
complexity of some sequences is unstable. For example, if a small
number of changes to a sequence greatly reduce its linear
complexity, then the resulting key stream is cryptographically weak.
Ding, Xiao and Shan noticed this problem first in their book
\cite{Ding}, and proposed the weight complexity and sphere
complexity. Stamp and Martin \cite{Stamp} introduced $k$-error
linear complexity, which is similar to the sphere complexity, and
put forward the concept of $k$-error linear complexity profile.
Specifically, suppose that $s$ is a sequence  with period $N$. For
any $k(0\le k\le N)$, $k$-error linear complexity of $s$, denoted as
$L_k(s)$,  is defined as the smallest linear complexity that can be
obtained when any $k$ or fewer bits of the sequence are changed
within one period.

One important result, proved by Kurosawa et al.  \cite{Kurosawa} is
that the minimum number $k$ for which the $k$-error linear
complexity of a $2^n$-periodic binary sequence $s$ is strictly less
than a linear complexity $L(s)$ of $s$ is determined by
$k_{\min}=2^{W_H(2^n-L(s))}$, where $W_H(a)$ denotes the Hamming
weight of the binary representation of an integer $a$. Also Rueppel
\cite{Rueppel} derived  the number $N(L)$ of $2^n$-periodic binary
sequences with given linear complexity $L, 0\le L \le 2^n$.

For $k=1,2$, Meidl \cite{Meidl2005} characterized the complete
counting functions on the $k$-error linear complexity of
$2^n$-periodic binary sequences having maximal possible linear
complexity $2^n$. For $k=2,3$, Zhu and Qi \cite{Zhu} further showed
the complete counting functions on the $k$-error linear complexity
of $2^n$-periodic binary sequences with linear complexity $2^n-1$.
By using algebraic and combinatorial methods, Fu et al. \cite{Fu}
studied the linear complexity and the $1$-error linear complexity
for $2^n$-periodic binary sequences, and characterized the
 $2^n$-periodic binary sequences with given 1-error linear
 complexity and derived the counting function for the 1-error
 linear complexity for $2^n$-periodic binary sequences.

By investigating sequences with linear complexity $2^n$ or linear
complexity less than $2^n$ together, Kavuluru
\cite{Kavuluru2008,Kavuluru} characterized $2^n$-periodic binary
sequences with fixed 2-error or 3-error linear complexity,  and
obtained the counting functions for the number of $2^n$-periodic
binary sequences with given $k$-error linear complexity for $k = 2$
and 3. These results are important progress on the $k$-error linear
complexity. Unfortunately, the results in
\cite{Kavuluru2008,Kavuluru} on the $3$-error linear complexity are
 not completely correct, as pointed out in \cite{Zhou}.

In current literature \cite{Meidl2005,Zhu,Kavuluru2008,Kavuluru},
sequences $s$  with $L_k(s)=c$ are directly investigated. In
contrast with that, we will study the $k$-error linear complexity by
proposing a new approach. Let $S=\{s | L(s)=c\}, E=\{e | W_H(e)\le
k\}, SE=\{s+e | s\in S, e\in E\}$, where $s$ is a sequence with
linear complexity $c$, and $e$ is an error sequence with $W_H(e)\le
k$. With the sieve method of combinatorics, we sieve sequences $s+e$
with $L_k(s+e)=c$ in $SE$.

First we investigate sequences with  linear complexity $2^n$,  and
sequences with linear complexity less than $2^n$, separately.  It is
observed that for sequences with  linear complexity $2^n$, the
$k$-error linear complexity is equal to  $(k+1)$-error linear
complexity, when $k$ is odd. For sequences with  linear complexity
less than $2^n$, the $k$-error linear complexity is equal to
$(k+1)$-error linear complexity, when $k$ is even. Then we
investigate the $3$-error linear complexity in two cases and this
reduces the complexity of this problem. Finally, by combining the
results of two cases, we obtain the complete counting functions for
the number of $2^n$-periodic binary sequences with  $3$-error linear
complexity.

The contribution of this paper can be summarized as follows. i) As
the results in \cite{Kavuluru2008,Kavuluru} on the $3$-error linear
complexity are  not completely correct,  the correct results are
given here. ii) A new approach is proposed for the $k$-error linear
complexity problem, which can decompose this problem into two sub
problems with less complexity. iii) Generally, the complete counting
functions for the number of $2^n$-periodic binary sequences with
given $k$-error linear complexity for $k>4$ can be obtained using a
similar approach.

\section{Preliminaries}

In this section we give some preliminary results which will be used in the sequel.

We will consider sequences over $GF(q)$, which is the finite field
of order $q$. Let $x=(x_1,x_2,\cdots,x_n)$ and
$y=(y_1,y_2,\cdots,y_n)$ be vectors over $GF(q)$. Then define
$x+y=(x_1+y_1,x_2+y_2,\cdots,x_n+y_n)$.

When $n=2m$, we define
$Left(x)=(x_1,x_2,\cdots,x_m)$ and $Right(x)=(x_{m+1},x_{m+2},\cdots,x_{2m})$.

The Hamming weight of an $N$-periodic sequence $s$ is defined as the
number of   nonzero elements in per period of $s$, denoted by
$W(s)$. Let $s^N$ be one period of $s$. If $N=2^n$, $s^N$ is also
denoted as $s^{(n)}$. Obviously, $W(s^{(n)})=W(s^N)=W(s)$. $supp(s)$
is defined as a set of the positions with  nonzero elements in per
period of $s$.

The generating function of a sequence $s=\{s_0, s_1, s_2, s_3,
\cdots, \}$  is defined by $$s(x)=s_0+ s_1x+ s_2x^2+ s_3x^3+
\cdots=\sum\limits^\infty_{i=0}s_ix^i$$

The generating function of a finite sequence $s^N=\{s_0, s_1, s_2,
 \cdots, s_{N-1},\}$ is defined by $s^N(x)=s_0+ s_1x+ s_2x^2+
\cdots+s_{N-1}x^{N-1}$. If $s$ is a periodic sequence with the first
period $s^N$, then,
\begin{eqnarray}
s(x) &=& s^N(x)(1+ x^N+ x^{2N}+ \cdots)=\frac{s^N(x)}{1-x^N}\notag\\
&=&\frac{s^N(x)/\gcd(s^N(x),1-x^N)}{(1-x^N)/\gcd(s^N(x),1-x^N)}\notag\\
&=&\frac{g(x)}{f_s(x)}\label{formula01}
\end{eqnarray}
where $f_s(x)=(1-x^N)/\gcd(s^N(x),1-x^N),
g(x)=s^N(x)/\gcd(s^N(x),1-x^N)$.

Obviously, $\gcd(g(x),f_s(x))=1, \deg(g(x)<\deg(f_s(x)))$. $f_s(x)$
is called  the minimal polynomial of $s$, and the degree of $f_s(x)$
is called the linear complexity of $s$, that is $\deg(f_s(x))=L(s)$.

Suppose that $N=2^n$ and $GF(q)=GF(2)$, then
$1-x^N=1-x^{2^n}=(1-x)^{2^n}=(1-x)^N$. Thus for binary sequences
with period $2^n$, its linear complexity is equal to the degree of
factor $(1-x)$ in $s^N(x)$.

The following three lemmas are  well known results on $2^n$-periodic
binary sequences.

\noindent {\bf Lemma  2.1} Suppose that s is a binary sequence with
period $N=2^n$, then $L(s)=N$ if and only if the Hamming weight of a
period of the sequence is odd.

If an element one is removed from a sequence whose Hamming weight is
odd, the Hamming weight of the sequence will be changed to even, so
the main concern hereinafter is about sequences whose Hamming weight
are even.

\noindent {\bf Lemma 2.2}  Let $s_1$ and $s_2$ be two binary sequences
with period $N=2^n$. If $L(s_1)\ne L(s_2)$, then
$L(s_1+s_2)=\max\{L(s_1),L(s_2)\} $; otherwise if $L(s_1)= L(s_2)$,
then $L(s_1+s_2)<L(s_1)$.

Suppose that the linear complexity of s can decline when at least
$k$ elements of s are changed. By Lemma 2.2, the linear complexity
of the binary sequence, in which elements at exactly those $k$
positions are all nonzero, must be L(s). Therefore, for the
computation of $k$-error linear complexity, we only need to find the
binary sequence whose Hamming weight is minimum and its linear
complexity is L(s).

\noindent {\bf Lemma  2.3} Let $E_i$ be a $2^n$-periodic sequence
with one nonzero element at position $i$ and 0 elsewhere in each
period, $0\le i<2^n$. If $j-i=2^r(1+2a), a\ge0, 0\le i<j<2^n,
r\ge0$, then $L(E_i +E_j)=2^n-2^r$.

\

\section{Counting functions with
the $k$-error linear complexity}

For $2^n$-periodic binary sequences with  linear complexity less than $2^n$,
 the change of one
bit per period results in a sequence with odd number of nonzero bits
per period, which has again linear complexity $2^n$. In this
section, we thus first focus on the $2$-error linear complexity.

Further more, in order to derive the  counting functions
on the $3$-error linear complexity of $2^n$-periodic binary
sequences with linear complexity less than $2^n$, we only need to investigate the $2$-error linear complexity of $2^n$-periodic binary
sequences with linear complexity less than $2^n$.

Second, for $k=3,4$, the complete counting functions on the
$k$-error linear complexity of $2^n$-periodic binary sequences with
linear complexity $2^n$ are presented. Third,  for $k=4,5$, the
complete counting functions on the $k$-error linear complexity of
$2^n$-periodic binary sequences with linear complexity less than
$2^n$ are derived.

Given a $2^n$-periodic binary sequence s, its linear complexity L(s)
can be  determined by the Games-Chan algorithm \cite{Games}. Based
on Games-Chan algorithm, the following  Lemma  3.1 is given in
\cite{Meidl2005}.

\noindent {\bf Lemma  3.1} Suppose that s is a binary sequence with
first period $s^{(n)}=\{s_0,s_1,s_2,\cdots, s_{2^n-1}\}$, a mapping
$\varphi_n$ from $F^{2^n}_2$ to $F^{2^{n-1}}_2$ is given by
\begin{eqnarray*}&&\varphi_n(s^{(n)})\\
&=&\varphi_n((s_0,s_1,s_2,\cdots,
s_{2^n-1}))\\
&=&(s_0+s_{2^{n-1}},s_1+s_{2^{n-1}+1},\cdots,
s_{2^{n-1}-1}+s_{2^n-1}) \end{eqnarray*}

Let $W(\mathbf{\upsilon})$ denote the Hamming weight of a vector
$\mathbf{\upsilon}$. Then mapping $\varphi_n$ has the following
properties

1) $W(\varphi_n(s^{(n)}))\le W(s^{(n)})$;

2) If $n\ge2$ then  $W(\varphi_n(s^{(n)}))$ and $W(s^{(n)})$ are
either both odd or both even;

3) The set $$\varphi^{-1}_{n+1}(s^{(n)})=\{v\in
F^{2^{n+1}}_2|\varphi_{n+1}(v)=s^{(n)} \}$$ of the preimage of
$s^{(n)}$ has cardinality $2^{2^n}$.

Rueppel \cite{Rueppel} presented the following.

\noindent {\bf Lemma  3.2}  The number $N(L)$ of $2^n$-periodic
binary sequences with given linear complexity $L, 0\le L \le 2^n$,
is given by $N(L)=\left\{\begin{array}{l}
1, \ \ \ \ \ L=0\ \   \\
2^{L-1}, \ 1\le L\le 2^n
\end{array}\right.$\

\

It is known that the computation of $k$-error linear complexity can be converted
to finding error sequences with minimal Hamming weight. Hence  2-error linear complexity of $s^{(n)}$ is  the smallest
linear complexity that can be obtained when any $u^{(n)}$ with
$W(u^{(n)})=0$ or 2 is added to $s^{(n)}$. So, the main approach of this section and next section is as follows. Let $s^{(n)}$ be a   binary sequence
with  linear complexity $c$, $u^{(n)}$  a  binary sequence with $W(u^{(n)})\le k$. We derive the  counting functions
on the $k$-error linear complexity of $2^n$-periodic binary
sequences by investigating $s^{(n)}+u^{(n)}$.
Based on this idea, we first prove the following lemmas

\noindent {\bf Lemma  3.3}  1). If $s^{(n)}$ is a   binary sequence
with  linear complexity $c, 1\le c\le 2^{n-1}-3$, $c\ne 2^{n-1}-2^m,
2\le m<n-1$,  $u^{(n)}$ is a  binary sequence, and $W(u^{(n)})=0$ or 2. Then the 2-error linear complexity
of $s^{(n)}+u^{(n)}$ is still $c$.

2). If $s^{(n)}$ is a   binary sequence with  linear complexity
$c=2^{n-1}-2^m, 0\le m<n-1$, then there exists a binary sequence
$u^{(n)}$ with $W(u^{(n)})=2$, such that the 2-error linear
complexity of $s^{(n)}+u^{(n)}$ is less than $c$.

\begin{proof}\
Without loss of  generality, we suppose that  $v^{(n)}\ne u^{(n)}$,
and  $W(v^{(n)})=0$ or 2.

1). As $c\le 2^{n-1}-3$, we only need to consider the case
$L(u^{(n)}+v^{(n)})<2^{n-1}$. Thus
$Left(u^{(n)}+v^{(n)})=Right(u^{(n)}+v^{(n)})$ and
$W(Left(u^{(n)}+v^{(n)}))=2$.

 By Lemma 2.3, $L(u^{(n)}+v^{(n)})=2^{n-1}-2^m$, $0\le m<n-1$.

Thus $L(s^{(n)}+u^{(n)}+v^{(n)})\ge L(s^{(n)})$, so the 2-error
linear complexity of $s^{(n)}+u^{(n)}$ is  $c$.

2). As $s^{(n)}$ is a   binary sequence with  linear complexity
$c=2^{n-1}-2^m, 0\le m<n-1$, so the 2-error linear complexity of
$s^{(n)}+u^{(n)}$ must be less than $c$ when $L(u^{(n)}+v^{(n)})=c$.
\end{proof}\

\noindent {\bf Lemma  3.4} Suppose that $s^{(n)}$ and $t^{(n)}$ are
two different  binary sequences with  linear complexity $c, 1\le
c\le 2^{n-2}$, and   $u^{(n)}$ and $v^{(n)}$  are two different  binary
sequences, and
$W(u^{(n)})=0$ or 2, and $W(v^{(n)})=0$ or 2. Then $s^{(n)}+u^{(n)}\ne
t^{(n)}+v^{(n)}$.

\begin{proof}\ The following is obvious

$s^{(n)}+u^{(n)}\ne t^{(n)}+v^{(n)}$

$\Leftrightarrow$ $s^{(n)}+u^{(n)}+v^{(n)}\ne t^{(n)}$

$\Leftrightarrow$ $u^{(n)}+v^{(n)}\ne s^{(n)}+t^{(n)}$

Note that $s^{(n)}$ and $t^{(n)}$ are two different  binary
sequences with  linear complexity $c, 1\le c\le 2^{n-2}$, so the
linear complexity of $s^{(n)}+t^{(n)}$ is less than $2^{n-2}$, hence
one period of $s^{(n)}+t^{(n)}$ can be divided into 4 equal parts.

Suppose that $u^{(n)}+v^{(n)}= s^{(n)}+t^{(n)}$, then one period of $u^{(n)}+v^{(n)}$ can be divided into 4 equal parts.
 It follows that
the linear complexity of $u^{(n)}+v^{(n)}$ is  $2^{n-2}$, which
contradicts the fact that the linear complexity of $s^{(n)}+t^{(n)}$
is less than $2^{n-2}$.
\end{proof}\

Next we divide the 2-error linear complexity into three categories.
First consider the category of $2^{n-1}-2^{n-m}$.

\noindent {\bf Lemma  3.5}  Let  $N_2(2^{n-1}-2^{n-m})$ be the
number of $2^n$-periodic binary sequences with  linear complexity less than
$2^n$ and given 2-error linear complexity $2^{n-1}-2^{n-m}, n\ge2,
1<m\le n$. Then
$$N_2(2^{n-1}-2^{n-m})=(1+\left(\begin{array}{c}2^n\\2\end{array}\right)-3\times2^{n+m-3}) 2^{2^{n-1}-2^{n-m}-1}
$$
\begin{proof}\
Suppose that $s^{(n)}$ is a   binary sequence with  linear
complexity $2^{n-1}-2^{n-m}$, and $u^{(n)}$ is a binary sequence
with  $W(u^{(n)})=2$. By Lemma  3.3, there exists a binary sequence $v^{(n)}$  with
 $W(v^{(n)})=2$, such that
$L(u^{(n)}+v^{(n)})=2^{n-1}-2^{n-m}$. So the 2-error linear
complexity of $u^{(n)}+s^{(n)}$ is less than $2^{n-1}-2^{n-m}$.

Suppose that $u^{(n)}$ is a binary sequence with linear complexity
$2^{n}$ and $W(u^{(n)})=2$, and there exist 2 nonzero elements whose
distance is  $2^{n-m}(2k+1)$ or $2^{n-1}$, with $k$ being an integer. It is
easy to verify that there exists a binary sequence $v^{(n)}$ with
 $W(v^{(n)})=2$, such that
$L(u^{(n)}+v^{(n)})=2^{n-1}-2^{n-m}$. So the 2-error linear
complexity of $u^{(n)}+s^{(n)}$ is less than $2^{n-1}-2^{n-m}$.

Let us divide one period of $u^{(n)}$   into $2^{n-m}$ subsequences
of form $\{a,a+2^{n-m}, a+2^{n-m+1},\cdots,
a+(2^m-1)\times2^{n-m}\}$.

If  2 nonzero elements of $u^{(n)}$ are in the same subsequence,
then the number of these $u^{(n)}$ can be given by
$$C1=2^{n-m}\times\left(\begin{array}{c}2^{m}\\2\end{array}\right)\times2^{m}.$$

 Suppose that  2 nonzero elements of
$u^{(n)}$ are in the same subsequence, and the distance of the 2
nonzero elements is not $2^{n-m}(2k+1)$, then the number of these
$u^{(n)}$ can be given by
$2^{n-m+1}\times\left(\begin{array}{c}2^{m-1}\\2\end{array}\right).$
Of  these $u^{(n)}$, there are
$2^{n-m}\times2^{m-1}=2^{n-1}$
sequences, in each sequence the distance of the 2 nonzero elements is
$2^{n-1}$.

So, if  2 nonzero elements of
$u^{(n)}$ are in the same subsequence, and the distance of the 2
nonzero elements is neither $2^{n-m}(2k+1)$ nor $2^{n-1}$, then the
number of these $u^{(n)}$ can be given by
$C2=2^{n-m+1}\times\left(\begin{array}{c}2^{m-1}\\2\end{array}\right)-2^{n-1}.$

Suppose that $u^{(n)}$ is a binary sequence with  $W(u^{(n)})=2$, and there exist  2 nonzero elements
whose distance is a multiple of $2^{n-m+1}$. Then there exists  one
binary sequence $v^{(n)}$  with
$W(v^{(n)})=2$, such that $L(u^{(n)}+v^{(n)})=2^{n-1}-2^{n-r},
1<r<m$. Let $t^{(n)}=s^{(n)}+u^{(n)}+v^{(n)}$. Then
$L(t^{(n)})=L(s^{(n)})=2^{n-1}-2^{n-m}$ and
$s^{(n)}+u^{(n)}=t^{(n)}+v^{(n)}$.

By Lemma  3.2,  the number  of $2^n$-periodic binary sequences with
given linear complexity $2^{n-1}-2^{n-m}$ is
$2^{2^{n-1}-2^{n-m}-1}$. This leads to the following,

{\scriptsize
\begin{eqnarray*}&&N_2(2^{n-1}-2^{n-m})\\
&=&[ 1+\left(\begin{array}{c}2^n\\2\end{array}\right)-(C1-C2)-C2/2]2^{2^{n-1}-2^{n-m}-1}\\
 &=&[
1+\left(\begin{array}{c}2^n\\2\end{array}\right)-2^{n-m}\left(\begin{array}{c}2^m\\2\end{array}\right)+2^{n-m}\left(\begin{array}{c}2^{m-1}\\2\end{array}\right)-2^{n-2}]\\
&&\times 2^{2^{n-1}-2^{n-m}-1}\\
&=&(1+\left(\begin{array}{c}2^n\\2\end{array}\right)-3\times2^{n+m-3}) 2^{2^{n-1}-2^{n-m}-1}\end{eqnarray*}}
\end{proof}\

Next we consider the category of $2^{n-1}-2^{n-m}+x$.

\noindent {\bf Lemma  3.6}  Let  $N_2(2^{n-1}-2^{n-m}+x)$ be the
number of $2^n$-periodic binary sequences with  linear complexity
less than $2^n$ and given 2-error linear complexity
$2^{n-1}-2^{n-m}+x, n>3, 1<m<n-1, 0<x<2^{n-m-1}$. Then{\small
\begin{eqnarray*}&&N_2(2^{n-1}-2^{n-m}+x)\\
&=&[1+\left(\begin{array}{c}2^n\\2\end{array}\right)+2^{n-m}-2^{n+m-2}]2^{2^{n-1}-2^{n-m}+x-1}\end{eqnarray*}}
\begin{proof}\
Suppose that $s^{(n)}$ is a binary sequence with linear complexity
$2^{n-1}-2^{n-m}+x$, and $u^{(n)}$ is a binary sequence with
$W(u^{(n)})=2$. By Lemma  3.3,  the 2-error linear complexity of
$u^{(n)}+s^{(n)}$ is still $2^{n-1}-2^{n-m}+x$.  The number of these
$u^{(n)}$ can be given by
$\left(\begin{array}{c}2^{n}\\2\end{array}\right).$

Suppose that $u^{(n)}$ is a binary sequence with  $W(u^{(n)})=2$,
and there exist  2 nonzero elements whose distance is
$2^{n-r}(1+2a), 1<r\le m, a\ge0$. Then there exists one binary
sequence $v^{(n)}$  with  $W(v^{(n)})=2$, such that
$L(u^{(n)}+v^{(n)})=2^{n-1}-2^{n-r}$. Let
$t^{(n)}=s^{(n)}+u^{(n)}+v^{(n)}$. Then
$L(t^{(n)})=L(s^{(n)})=2^{n-1}-2^{n-m}+x$ and
$s^{(n)}+u^{(n)}=t^{(n)}+v^{(n)}$.

Let us divide one period of $u^{(n)}$   into $2^{n-m}$ subsequences
of form $\{a,a+2^{n-m}, a+2^{n-m+1},\cdots,
a+(2^m-1)\times2^{n-m}\}$.

If  2 nonzero elements of $u^{(n)}$ are in the same subsequence, and
their distance is
 $2^{n-1}$,  then there
exist $2^{m-1}-1$ binary sequences $v^{(n)}$ with $W(v^{(n)})=2$,
such that $L(u^{(n)}+v^{(n)})=2^{n-1}-2^{n-r}, 1<r\le m$. Let
$t^{(n)}=s^{(n)}+u^{(n)}+v^{(n)}$. Then
$s^{(n)}+u^{(n)}=t^{(n)}+v^{(n)}$. The number of these $u^{(n)}$ can
be given by $D1=2^{n-m}\times2^{m-1}=2^{n-1}.$

Suppose that 2 nonzero elements of $u^{(n)}$ are in the same
subsequence, and their distance is not  $2^{n-1}$.
 Then there exist  one binary sequence $v^{(n)}$,  with $W(v^{(n)})=2$, such that
$L(u^{(n)}+v^{(n)})=2^{n-1}-2^{n-r}, 1<r\le m$.  The number of these
$u^{(n)}$ can be given by
$$D2=2^{n-m}[\left(\begin{array}{c}2^{m}\\2\end{array}\right)-2^{m-1}]$$

By Lemma  3.2,  the number  of $2^n$-periodic binary sequences with
given linear complexity $2^{n-1}-2^{n-m}+x$ is
$2^{2^{n-1}-2^{n-m}+x-1}$. This will derive the following, {\small
\begin{eqnarray*}&&N_2(2^{n-1}-2^{n-m}+x)\\
&=&[1+ \left(\begin{array}{c}2^n\\2\end{array}\right)- \frac{2^{m-1}-1}{2^{m-1}}\times D1-\frac{1}{2}\times D2]\\
&&\ \ \ \ 2^{2^{n-1}-2^{n-m}+x-1}\\
&=&\{1+ \left(\begin{array}{c}2^n\\2\end{array}\right)- \frac{2^{m-1}-1}{2^{m-1}}\times 2^{n-1}\\
&&-2^{n-m-1}[\left(\begin{array}{c}2^m\\2\end{array}\right)-2^{m-1}]\}2^{2^{n-1}-2^{n-m}+x-1}\\
&=&\{1+\left(\begin{array}{c}2^n\\2\end{array}\right)-(2^{m-1}-1)\times2^{n-m}\\
&&-2^{n-m-1}[\left(\begin{array}{c}2^{m}\\2\end{array}\right)-2^{m-1}]\}2^{2^{n-1}-2^{n-m}+x-1}\\
&=&[1+\left(\begin{array}{c}2^n\\2\end{array}\right)+2^{n-m}-2^{n+m-2}]2^{2^{n-1}-2^{n-m}+x-1}
\end{eqnarray*}}
\end{proof}\

\

Finally we consider the simplest category, that is $1\le c\le
2^{r-2}-1$.

\noindent {\bf Lemma  3.7}  Let $L(r,c)=2^n-2^r+c, 3\le r\le n, 1\le
c\le 2^{r-2}-1$, and $N_2(L(r,c))$ be the number of $2^n$-periodic
binary sequences with  linear complexity less than $2^n$ and given
2-error linear complexity $L(r,c)$. Then
$$N_2(L)=\left\{\begin{array}{l}
1+\left(\begin{array}{c}2^n\\2\end{array}\right), \ \ \ \ \ \ \  \ L=0\ \   \\
2^{L-1}(1+\left(\begin{array}{c}2^r\\2\end{array}\right)), \
L=L(r,c)
\end{array}\right.$$
\begin{proof}\
Suppose that $s$ is a binary sequence with first period
$s^{(n)}=\{s_0,s_1,s_2,\cdots, s_{2^n-1}\}$, and $L(s)<2^n$. By
Games-Chan algorithm, $Left(s^{(t)})\ne Right(s^{(t)}), 1\le t\le
n$,
 where
$s^{(t)}=\varphi_{t+1}\cdots\varphi_{n}(s^{(n)})$.

First consider the  case of $W(s^{(n)})=0$. There is only one binary
sequence of this kind.

Consider the  case of $W(s^{(n)})=2$. There is 2 nonzero bits in
$\{s_0,s_1,\cdots, s_{2^n-1}\}$, thus there are
$\left(\begin{array}{c}2^n\\2\end{array}\right)$ binary sequences of
this kind.

So $N_2(0)=1+\left(\begin{array}{c}2^n\\2\end{array}\right)$.

Consider $L(r,c)=2^n-2^r+c$, $3\le r\le n, 1\le c\le 2^{r-2}-1$.
Suppose that $s^{(n)}$ is a binary sequence with
$L(s^{(n)})=L(r,c)$. Note that
$L(r,c)=2^n-2^r+c=2^{n-1}+\cdots+2^r+c$.  By Games-Chan algorithm,
$Left(s^{(r)})= Right(s^{(r)})$, and $L(s^{(r)})=c$.

It is known that the number of  binary sequences $t^{(r)}$ with
$W(t^{(r)})=0$ or 2  is
$1+\left(\begin{array}{c}2^r\\2\end{array}\right)$.

By Lemma 3.3, the 2-error linear complexity of $s^{(r)}+t^{(r)}$ is
$c$.

By Lemma 3.2 and Lemma 3.4, the number of  binary sequences
$s^{(r)}+t^{(r)}$ is $2^{c-1}\times
(1+\left(\begin{array}{c}2^r\\2\end{array}\right))$

By Lemma 3.1, there are $2^{2^{n-1}+\cdots+2^r}=2^{2^{n}-2^r}$
binary sequences $s^{(n)}+t^{(n)}$,  such that
$s^{(r)}+t^{(r)}=\varphi_{r+1}\cdots\varphi_{n}(s^{(n)}+t^{(n)})$,
$t^{(r)}=\varphi_{r+1}\cdots\varphi_{n}(t^{(n)})$ and
$W(t^{(n)})=W(t^{(r)})$.

 Thus  the 2-error linear complexity of
$s^{(n)}+t^{(n)}$ is
$$2^{n-1}+\cdots+2^r+L_2(s^{(r)}+t^{(r)})=2^{n}-2^r+c=L(r,c).$$

Therefore, $N_2(L(r,c))=2^{2^{n}-2^r}\times 2^{c-1}\times
(1+\left(\begin{array}{c}2^r\\2\end{array}\right))=2^{L(r,c)-1}(1+\left(\begin{array}{c}2^r\\2\end{array}\right))$
\end{proof}\

Based on the results above, we have the following theorem.

 \noindent {\bf Theorem  3.1}  Let $L(r,c)=2^n-2^r+c$,  $2\le r\le
n, 1\le c\le 2^{r-1}-1$, and $N_2(L(r,c))$ be the number of
$2^n$-periodic binary sequences with  linear complexity less than
$2^n$ and given 2-error linear complexity $L(r,c)$.
 Then

%{\scriptsize%\small
{\scriptsize

$N_2(L)=\left\{\begin{array}{l}
\left(\begin{array}{c}2^{n}\\2\end{array}\right)+1, \ \ \ \ \ \ \ \ \ \ L=0\ \   \\
2^{L-1}(\left(\begin{array}{c}2^{r}\\2\end{array}\right)+1), \ L=L(r,c), 1\le c\le 2^{r-2}-1, r>2\\
2^{L-1}(\left(\begin{array}{c}2^{r}\\2\end{array}\right)+1-3\times2^{r+m-3}),  \\
 \ \ \ \ \  \ \ \ \ \ \ \ \ \ L=L(r,c),  c= 2^{r-1}-2^{r-m},  1<m\le r,r\ge2\\
2^{L-1}(\left(\begin{array}{c}2^{r}\\2\end{array}\right)+1+2^{r-m}-2^{r+m-2}),  \\
 \ \ \ \ \  \ \ \ \ \ \ \ \  L=L(r,c),  c= 2^{r-1}-2^{r-m}+x, \\
\ \ \ \ \  \ \ \ \ \ \ \ \   1<m<r-1, 0<x<2^{r-m-1}, r>3\\
0, \ \ \ \ \ \ \  \ \ \ \ \ \ \ \ \ \  \ \ \ \ \ \ \ \ \
\mbox{others}
\end{array}\right.$}
\begin{proof}\

By Lemma 3.7, we now only need to consider the case of $3\le r\le n,
2^{r-2}\le c\le 2^{r-1}-1$.

By Lemma 3.1 and Lemma 3.5,
$$N_2(L(r,c))=2^{L(r,c)-1}(\left(\begin{array}{c}2^{r}\\2\end{array}\right)+1-3\times2^{r+m-3})$$
for $3\le r\le n,   c= 2^{r-1}-2^{r-m}, 1<m\le r$

By Lemma 3.1 and Lemma 3.6,
$$N_2(L(r,c))=2^{L(r,c)-1}(\left(\begin{array}{c}2^{r}\\2\end{array}\right)+1+2^{r-m}-2^{r+m-2})$$
for $4\le r\le n,   c= 2^{r-1}-2^{r-m}+x,1<m<r-1,0<x<2^{r-m-1}$

This completes the proof.
\end{proof}\

\

Now we give an example to illustrate Theorem 3.1.

For $n = 4$, the number of $2^n$-periodic binary sequences with
 linear complexity less than $2^n$ is $count = 2^{2^4-1} = 2^{15}$.

$N_2(L(2,1))=2^{12}$

$N_2(L(3,1))=2^8[\left(\begin{array}{c}2^{3}\\2\end{array}\right)+1]=count\times\frac{29}{128}$.

$N_2(L(3,2))=count\times\frac{17}{64}$.

$N_2(L(3,3))=count\times\frac{5}{32}$.

$N_2(0)=N_2(L(4,1))=\left(\begin{array}{c}2^{4}\\2\end{array}\right)+1=121$.

$N_2(L(4,2))=2\times121=242$.

$N_2(L(4,3))=4\times121=484$.

 $N_2(L(4,4))=776$.\ \ \
 $N_2(L(4,5))=1744$.

$N_2(L(4,6))=2336$.\ \ $N_2(L(4,7))=1600$.

It is easy to verify that the number of all these sequences is
$2^{15}$. These results are also checked by computer.

\

Notice that for $2^n$-periodic binary sequences with  linear complexity less than $2^n$,
 the change of three
bits per period results in a sequence with odd number of nonzero
bits per period, which has again linear complexity $2^n$. So from Theorem  3.1, we also know the counting functions on
the $3$-error linear complexity for $2^n$-periodic binary sequences with  linear complexity less than $2^n$.

Similarly, we can have the following theorem.

 \noindent {\bf Theorem  3.2}  Let $L(r,c)=2^n-2^r+c$, or $2^n-2^3+1$, $4\le r\le
n, 1\le c\le 2^{r-1}-1$, and $N_3(L(r,c))$ be the number of
$2^n$-periodic binary sequences with  linear complexity $2^n$ and
given 3-error linear complexity $L(r,c)$. Let {\scriptsize
\begin{eqnarray*}&&f(n,m)\\
&=&
\left(\begin{array}{c}2^n\\3\end{array}\right)-2^{n-m}\left(\begin{array}{c}2^m\\3\end{array}\right)-\left(\begin{array}{c}2^{n-m}\\2\end{array}\right)\left(\begin{array}{c}2^m\\2\end{array}\right)2^{m+1}\\
&&+\left(\begin{array}{c}2^{n-m}\\2\end{array}\right)\times2^{2m}(2^{m-2}-1)+2^{n-m-1}\times\left(\begin{array}{c}2^{m-1}\\3\end{array}\right)\\
&&-2^{n-2}\times(2^{m-2}-1)\\
\ \\
&&g(n,m)\\
&=&
\left(\begin{array}{c}2^n\\3\end{array}\right)-(2^{m-2}-1)\times2^{n+1}\\
&&-(2^{m-1}-1)\times\left(\begin{array}{c}2^{n-m}\\2\end{array}\right)\times2^{m+1}\\
&&-3\times2^{n-m-2}[\left(\begin{array}{c}2^m\\3\end{array}\right)-4\left(\begin{array}{c}2^{m-1}\\2\end{array}\right)]\\
&&-\left(\begin{array}{c}2^{n-m}\\2\end{array}\right)\times[\left(\begin{array}{c}2^m\\2\end{array}\right)-2^{m-1}]\times2^m\end{eqnarray*}}

 Then

{\scriptsize $N_3(L)=\left\{\begin{array}{l}
\left(\begin{array}{c}2^{n}\\3\end{array}\right)+2^n,  \ \ \  \ L=0\ \   \\
2^{L(r,c)-1}(\left(\begin{array}{c}2^{r}\\3\end{array}\right)+2^r),\\
 \ \ \ \ \  \ \ \  L=L(r,c), 1\le c\le 2^{r-2}-1, r>2\\
2^{L(r,c)-1}f(r,m), \\
\ \ \ \ \ \ \ \ L=L(r,c),  c= 2^{r-1}-2^{r-m},1<m\le r, r>3\\
2^{L(r,c)-1}g(r,m),\\
 \ \ \ \ \  \ \ \  L=L(r,c),  c= 2^{r-1}-2^{r-m}+x,\\
 \ \ \ \ \  \ \ \    1<m<r-1,0<x<2^{r-m-1},r>3\\
0, \ \ \ \ \ \ \  \ \ \ \ \ \ \ \ \ \  \ \ \ \  \ \mbox{others}
\end{array}\right.$}

\

Based on Theorem  3.1 and Theorem  3.2, the counting functions  for
the number of $2^n$-periodic binary sequences with fixed 3-error
linear complexity can be easily derived as follows.

 \noindent {\bf Theorem  3.3}  Let $L(r,c)=2^n-2^r+c$,  $4\le r\le
n, 1\le c\le 2^{r-1}-1$, and $N_3(L(r,c))$ be the number of
$2^n$-periodic binary sequences with  3-error linear complexity
$L(r,c)$.
 Then

{\scriptsize

$N_3(L)=\left\{\begin{array}{l}
\left(\begin{array}{c}2^{n}\\3\end{array}\right)+\left(\begin{array}{c}2^{n}\\2\end{array}\right)+2^n+1, \ \ \ \  \ \ \ L=0\ \   \\
2^{L-1}(\left(\begin{array}{c}2^{r}\\3\end{array}\right)+\left(\begin{array}{c}2^{r}\\2\end{array}\right)+2^r+1), \\
 \ \ \ \ \  \ \ \ \ \ \ \ \ \ L=L(r,c), 1\le c\le 2^{r-2}-1, r>3\\
2^{L-1}(\left(\begin{array}{c}2^{r}\\2\end{array}\right)+1-3\times2^{r+m-3}+f(r,m)),  \\
 \ \ \ \ \  \ \ \ \ \ \ \ \ \ L=L(r,c),  c= 2^{r-1}-2^{r-m},  1<m\le r,r>3\\
2^{L-1}(\left(\begin{array}{c}2^{r}\\2\end{array}\right)+1+2^{r-m}-2^{r+m-2}+g(r,m)),  \\
 \ \ \ \ \  \ \ \ \ \ \ \ \  L=L(r,c),  c= 2^{r-1}-2^{r-m}+x, \\
\ \ \ \ \  \ \ \ \ \ \ \ \   1<m<r-1, 0<x<2^{r-m-1}, r>3\\
0, \ \ \ \ \ \ \  \ \ \ \ \ \ \ \ \ \  \ \ \ \ \ \ \ \ \ \ \  \ \ \
\mbox{others}
\end{array}\right.$} where $f(r,m)$ and $g(r,m)$ are defined in Theorem
3.2.

\

Let $L(r,c)=2^n-2^r+c$, $3\le r\le n, 1\le c\le 2^{r-1}-1$. By
dividing the 4-error linear complexity into six categories:
$c=2^{r-2}-2^{r-m}$,  $c=2^{r-2}-2^{r-m}+x$,  $c=2^{r-1}-2^{r-m}$,
$c=2^{r-1}-(2^{r-m}+2^{r-j})$, $c=2^{r-1}-(2^{r-m}+2^{r-j})+x$, and
$1\le c\le 2^{r-3}-1$, we finally got the  counting functions  for
the number of $2^n$-periodic binary sequences with linear complexity
less than $2^n$ and fixed 4-error linear complexity. As a
consequence of the result, the complete counting functions on the
$4$-error linear complexity of $2^n$-periodic  binary sequences
(with linear complexity $2^n$ or less than $2^n$) are obvious.

Here only  the cases of $c=2^{r-1}-2^{r-m}$ and
$c=2^{r-1}-(2^{r-m}+2^{r-j})$ are presented. The results about other
cases are omitted.

\noindent {\bf Lemma  3.8}  Let $N_4(2^{n-1}-2^{n-m})$ be the number
of $2^n$-periodic binary sequences with linear complexity less than
$2^n$ and given 4-error linear complexity $2^{n-1}-2^{n-m}, 2\le
m\le n$. Then {\scriptsize
\begin{eqnarray*}&&N_4(2^{n-1}-2^{n-m})\\
&=&[\left(\begin{array}{c}2^n\\4\end{array}\right)-E1+E2/4-E3+E4/2-E5+E6/4-E7+E8/8]\\
&&\ \ \times2^{2^{n-1}-2^{n-m}-1}\end{eqnarray*}} where {\scriptsize
$$E1=\left(\begin{array}{c}2^{n-m}\\2\end{array}\right)\times\left(\begin{array}{c}2^{m}\\2\end{array}\right)
\times\left(\begin{array}{c}2^{m}\\2\end{array}\right)$$

\begin{eqnarray*}
&&E2=4\times\left(\begin{array}{c}2^{n-m}\\2\end{array}\right)\times\left(\begin{array}{c}2^{m-1}\\2\end{array}\right)\times\left(\begin{array}{c}2^{m-1}\\2\end{array}\right)\\
&& \ \ \ \ \ \ \ -\left(\begin{array}{c}2^{n-m}\\2\end{array}\right)\times[2^{2m-2}+2^{m+1}(\left(\begin{array}{c}2^{m-1}\\2\end{array}\right)\\
&& \ \ \ \ \ \ \ -2^{m-2})]
\end{eqnarray*}

$$E3=\left(\begin{array}{c}2^{n-m}\\3\end{array}\right)\times\left(\begin{array}{c}3\\1\end{array}\right)\times\left(\begin{array}{c}2^{m}\\2\end{array}\right)\times2^m\times2^m$$

\begin{eqnarray*}
&&E4=\left(\begin{array}{c}2^{n-m}\\3\end{array}\right)\left(\begin{array}{c}3\\1\end{array}\right)\left(\begin{array}{c}2^{m-1}\\2\end{array}\right)\times2^{2m+1} \\
&&\ \ \ \ \ \ \ \
-\left(\begin{array}{c}2^{n-m}\\3\end{array}\right)\left(\begin{array}{c}3\\1\end{array}\right)\times2^{3m-1}\end{eqnarray*}

$$ E5=\left(\begin{array}{c}2^{n-m}\\2\end{array}\right)\left(\begin{array}{c}2\\1\end{array}\right)\left(\begin{array}{c}2^m\\3\end{array}\right)\times2^m $$

\begin{eqnarray*}
&&E6=2^{m+2}\times\left(\begin{array}{c}2^{n-m}\\2\end{array}\right)\times\left(\begin{array}{c}2^{m-1}\\3\end{array}\right)\\
&&\ \ \ \ \ \ \ \
-\left(\begin{array}{c}2^{n-m}\\2\end{array}\right)\times(2^{m-1}-2)\times2^{2m}\end{eqnarray*}

$$E7=2^{n-m}\times\left(\begin{array}{c}2^{m}\\4\end{array}\right)$$

\begin{eqnarray*}
&&E8=2^{n-m+1}\times\left(\begin{array}{c}2^{m-1}\\4\end{array}\right)-[2^{n-m+1}\times2^{m-2}\\
&&\ \ \ \ \ \ \ \
\times\left(\begin{array}{c}2^{m-1}-2\\2\end{array}\right)-2^{n-m+1}\times\left(\begin{array}{c}2^{m-2}\\2\end{array}\right)]
\end{eqnarray*}}

By Lemma 3.8, for $n=5,m=5$, $N_4(15)=4587520$, which is checked by
computer.

\noindent {\bf Lemma  3.9}  Let $N_4(2^{n-1}-(2^{n-m}+2^{n-j}))$ be
the number of $2^n$-periodic binary sequences with linear complexity
less than $2^n$ and given 4-error linear complexity
$2^{n-1}-(2^{n-m}+2^{n-j}), n>3, 2<m<j\le n$. Then {\scriptsize
\begin{eqnarray*}&&N_4(2^{n-1}-(2^{n-m}+2^{n-j}))\\
&=&[1+\left(\begin{array}{c}2^n\\2\end{array}\right)+\left(\begin{array}{c}2^n\\4\end{array}\right)-F4\\
&&-\sum\limits_{k=m+1}^{j-1}(\frac{2^{2m-3}-1}{2^{2m-3}}F6+\frac{2^{m-1}-1}{2^{m-1}}F7+F8/2)\\
&&-\frac{2^{m-2}-1}{2^{m-2}}F10-F11/2-F13-\frac{3}{4}F14-\frac{2^{m-1}-1}{2^{m-1}}F17\\
&&-\frac{3}{4}F18-\frac{2^{2m-4}-1}{2^{2m-4}}F19-F22/2-\frac{2^{m-2}-1}{2^{m-2}}F23\\
&&-F25-F26-\frac{7}{8}F27]\times2^{2^{n-1}-(2^{n-m}+2^{n-j})-1}
\end{eqnarray*}}
where {\small $$
F4=2^{n+2m+j-6}+2^{n+m-4}+2^{n+j-4}+3\times2^{n+m+j-4}$$
$$F6=2^{n+k-4},F7=3\times2^{n+m+k-4},F8=2^{n+2m+k-6}
$$
$$F10=2^{n-1},F11=2^{n-m+1}\left(\begin{array}{c}2^{m-1}\\2\end{array}\right)-2^{n-1}$$
$$F13=\left(\begin{array}{c}2^{n-m+1}\\2\end{array}\right)\times\left(\begin{array}{c}2\\1\end{array}\right)\times2^{m-2}\times(2^{m-1}-2)\times2^{m-1}$$

$$F14=\left(\begin{array}{c}2^{n-m+1}\\2\end{array}\right)\times\frac{2^{m-1}-4}{3}\times(2^{m-1}-2)\times2^{2m-2}$$
$$F17=\left(\begin{array}{c}2^{n-m+1}\\2\end{array}\right)\times\left(\begin{array}{c}2\\1\end{array}\right)\times2^{m-2}\times[\left(\begin{array}{c}2^{m-1}\\2\end{array}\right)-2^{m-2}]$$
$$F18=\left(\begin{array}{c}2^{n-m+1}\\2\end{array}\right)\times[\left(\begin{array}{c}2^{m-1}\\2\end{array}\right)-2^{m-2}]^2
$$
$F19=\left(\begin{array}{c}2^{n-m+1}\\2\end{array}\right)\times\left(\begin{array}{c}2^{m-1}\\2\end{array}\right)^2-2^{n+m-4}
-\sum\limits_{k=m+1}^j 2^{n+k-4}-F17-F18$
$$F22=\left(\begin{array}{c}2^{n-m+1}\\3\end{array}\right)\left(\begin{array}{c}3\\1\end{array}\right)\times(\left(\begin{array}{c}2^{m-1}\\2\end{array}\right)-2^{m-2})\times(2^{m-1})^2
$$
$F23=\left(\begin{array}{c}2^{n-m+1}\\3\end{array}\right)\times\left(\begin{array}{c}3\\1\end{array}\right)\times\left(\begin{array}{c}2^{m-1}\\2\end{array}\right)\times(2^{m-1})^2-3\sum
\limits_{k=m+1}^j 2^{n+m+k-4}-F22 $
$$F25=2^{n-m+1}\times\left(\begin{array}{c}2^{m-2}\\2\end{array}\right)$$
$$F26=2^{n-m+1}\times2^{m-2}\times[\left(\begin{array}{c}2^{m-1}-2\\2\end{array}\right)-(2^{m-2}-1)]$$
$$F27=2^{n-m+1}\times\left(\begin{array}{c}2^{m-1}\\4\end{array}\right)-F25-F26$$}

By Lemma 3.9, for $n=5,m=4, j=5$, $N_4(13)=46845952$, which is
checked by computer.

\section{Conclusion}

 By using the sieve method of combinatorics, an approach to construct the complete counting functions
on the $k$-error linear complexity of $2^n$-periodic binary
sequences  was developed. The complete counting functions on the
$k$-error linear complexity of $2^n$-periodic binary sequences were
obtained for $k=3$ and 4.

Using the approach proposed, we can deal with the $k$-error linear
complexity distribution of sequences over $GF(q)$ with period $p^n$
or $2p^n$, where $p$ and $q$ are odd primes, and $q$ is a primitive
root modulo $p^2$.

 \section*{ Acknowledgment}
 The research was supported by
Zhejiang Natural Science Foundation(No.Y1100318, R1090138) and NSAF
(No. 10776077).

\end{document}